# A Novel and Reliable Deep Learning Web-Based Tool to Detect COVID-19 Infection from Chest CT-Scan

Abdolkarim Saeedi [1] · Maryam Saeedi [1] · Arash Maghsoudi [1]

*Abstract*

The corona virus is already spread around the world in many countries, and it has taken many lives. Furthermore, the world health organization (WHO) has announced that COVID-19 has reached the global epidemic stage. Early and reliable diagnosis using chest CT-scan can assist medical specialists in vital circumstances. In this work, we introduce a computer aided diagnosis (CAD) web service to detect COVID-19 online. One of the largest public chest CT-scan databases, containing 746 participants was used in this experiment. A number of well-known deep neural network architectures consisting of ResNet, Inception and MobileNet were inspected to find the most efficient model for the hybrid system. A combination of the Densely connected convolutional network (DenseNet) in order to reduce image dimensions and Nu-SVM as an anti-overfitting bottleneck was chosen to distinguish between COVID-19 and healthy controls. The proposed methodology achieved 90.80% recall, 89.76% precision and 90.61% accuracy. The method also yields an AUC of 95.05%. Ultimately a flask web service is made public through ngrok using the trained models to provide a RESTful COVID-19 detector, which takes only 39 milliseconds to process one image. The source code is also available at https://github.com/KiLJ4EdeN/COVID_WEB. Based on the findings, it can be inferred that it is feasible to use the proposed technique as an automated tool for diagnosis of COVID-19.

1.Introduction

The word "corona virus" is derived from the Latin word "corona" which implies crown or halo. The term refers to the appearance of the infectious form of the virus under electronic microscope.

Abdolkarim Saeedi
a-saeedi@srbiau.ac.ir

Maryam Saeedi
Maryam.saeedi33@gmail.com

Arash Maghsoudi
maghsoudi@srbiau.ac.ir

[1] Department of Biomedical Engineering, Science and Research Branch, Islamic Azad University, Tehran, Iran.

So far, six specimens of the coronavirus have been detected, and with the most recent case, the number has risen to seven. Corona virus disease (COVID-19) is a novel undiscovered type of coronavirus to humans. Reporting of the virus was first in Wuhan, China. Since then, it was spread rapidly and widely in China, and cases have been reported in many other countries. Fever, cough and difficulty breathing are usual signs. Symptoms tend to start about after five days, but can vary between two fourteen days. Human coronavirus is transmitted through infected droplets distributed by coughing or sneezing, or by contacting with contaminated skin and surfaces. On January 30[th], the World Health Organization (WHO) declared the 2019-20 coronavirus outbreak a Public Health Emergency of International Concern and a pandemic on March 11th 2020. According to WHO reports more than 3.23 million are affected by the coronavirus which yields in 228,394 deaths by the end of April 2020 [1]. The typical diagnostic approach is from a nasopharyngeal swab by real-time reverse transcription polymerase chain reaction (rRT-PCR). Due to the limited number of test kits and the amount of time to get the response, CT-scan based detection can be a promising tool. Deep learning in the field of automatic diagnosis of different disorders has made considerable improvements and it is able to become a fast, accurate and more accessible method to help medical professionals in critical conditions. Wang et Al [2] have analyzed 1119 CT-scan images using transfer learning. Their results demonstrate 79.3% accuracy. Zhao et Al [3], have provided a CT-scan public database. They have used a pre-trained deep learning conventional neural network (CNN) based on the chest X-ray images for discrimination between 247 COVID-19 patients and healthy controls which results in 84.7% accuracy. They used transfer learning and data augmentation to avoid overfitting. In another study, Wang et Al [4] have combined two publicly available datasets, a total 16,756 images where COVIDx dataset consists of 76 COVID-19 images from 53 patients and the other dataset is related to RSNA pneumonia detection challenge. They have developed a new neural network architecture; COVID-Net architecture in order to detect COVID-19 based on chest X-ray images. Their results revealed 92.4% accuracy on the COVIDx dataset. Xu et Al [5] experimented with 219 images from 110 patients and 224 influenza-A CT-scan images. They trained a 3-dimensional CNN to segment the infected areas aiming to differentiate between COVID-19 and influenza-A patients. Right now, as the models and methods for evaluating COVID-19 is progressing very quickly, the need to develop a fast and reliable diagnosis technique, with the ability to be used anywhere in the real-world applications seems necessary. In this paper, we propose a fast pipeline using a combination of deep and machine learning algorithms. The DenseNet121 presents remarkable results based on the Nu-SVM classifier. The introduced hybrid model is also compared to other common deep learning methods to assert comprehensive empirical evidence for the reliability of the proposed combination. The models are also hosted on an online web service, to be to perform live detection based on chest CT images. By providing the source code and a detailed explanation, we further insist on providing enough information for the community to keep on with the pandemic.

2. Materials and Methods
2.1 Database Description
The dataset used in this study, which is publicly available [6] consists of 349 patients with confirmed COVID and 397 healthy subjects. For processing purposes images were resized to (224, 224, 3), as suggested in the article [3] Images were also normalized for better compensation with learning method [7]. Figure 1 demonstrates two COVID and non COVID samples.

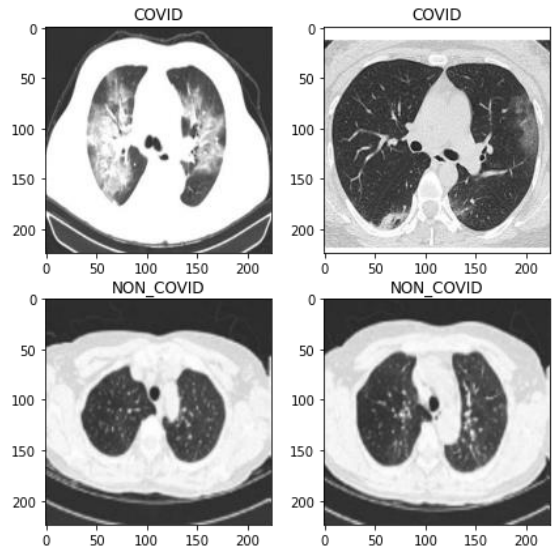

Figure 1. Examples of CT-scan images

2.2 DenseNet121
DenseNets are an extension of Convolutional neural networks. The main idea behind DenseNet is to reduce parameters, as well as having a boost to computational efficiency by using all of the outputs form the previous layers as inputs to the next ones. Also, this model has a less tendency to overfit [8] compared to other techniques such as ResNet [9] or FractalNet [10]. Figure 2 shows the architecture of this network. Also, equation 1 shows the formula for a Densely connected layer:

$$x_l = h_l([x_0, \ldots, x_{l-1}]) \qquad (1)$$

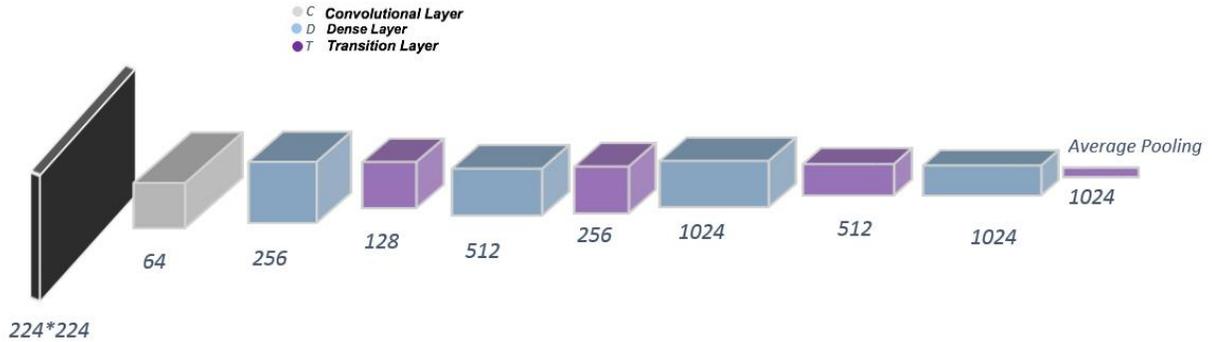

Figure 2. Architecture of Densest121

Where $x_l$ is the output of the current layer represented by $l$ with a non-linear transformation denoted by $h_l$. As it can be observed, the current layer reuses all the features from $0$ to $l-1$ layers. In this work, DenseNet121 was chosen due to less overfitting with respect to other available variations [8].

2.3 Inception
In order to design deeper networks and address the budget challenges, Inception neural networks were implemented through reducing the dimensionality via usage of $1*1$ stacked convolutions. The aim is to get multiple kernel sizes inside the network instead of sequentially stacking them, ordering each to work at the same stage. Initial version of inception architecture generally known as GoogLeNet which was introduced by Szegedy et Al [11]. Extra modifications were built on the early model by presenting batch normalization (inceptionV2) and factorization (inceptionV3) with aiming to reduce computational expanse and parameters by keeping the network performance stable [12].

2.4 ResNet50
ResNet (Residual Network) architecture was released in 2015 and won multiple competitions. With ResNet152 known as the deepest structure for some time. ResNet50 is a 50 layered deep learning neural network. The network is composed of five stages which each of these stages includes convolutional and identity blocks.

Main idea behind ResNet is to identity mapping layers that skips one or several layers. Therefore, in each stage, in addition to the results obtained from previous layer, the data itself is also fed to the next layer which makes deeper models trainable with an easier optimization process [9].

2.5 MobileNet
MobileNet is a fast and lightweight deep neural network which is appropriate for mobile and embedded vision applications [13]. This architecture employs depth-wise separable convolutions which uses a depth-wise operation followed by a point-wise convolution. The idea of depth-wise separable convolution is that instead of the usual convolution operation -where each filter is applied to all input features and then the output is provided by the combination of them- the process is comprised of two steps; once the filter has been applied to a feature map, the results are not combined immediately. In the second step a $1*1$ convolution layer performs the combination of feature maps yielded in the previous step. This method greatly reduces the number of parameters compared to a traditional network with identical depth.

2.6 Nu-SVM
SVM was first introduced by Vapnik et al in 1963 [14] as a linear classifier. This method uses geometrical parameters instead of statistical therefore it's a non-parametric classifier. In general, SVM is used for two or multi-class classification and regression problems. For N training samples each demonstrated with $D = \{x_i, y_i\}$, $x_i$ referred to as feature vector and $y_i$ is set of labels. The idea behind the SVM is to find the optimal hyperplane with maximum margin in order to differentiate between classes. If the samples are not linearly separable, non-linear kernel SVM projects the data into N-dimensional space and the decision boundary is determined in that space. The equation of the hyperplane is as follows:

$$w^T \varphi(x) + b = 0 \quad (2)$$

Where x represents each sample the on hyperplane, $w$ is a vector perpendicular to decision boundary, b is bias, $\varphi$ is kernel function for transformation of dataset to N-dimensional space. Nu-SVM is distinguished from conventional C-SVM by means of regularization parameter. C varies between zero to infinity and hardly can be estimated while $\nu$ operates between zero and one. Optimization function for minimizing the loss can be obtained by following equation:

$$\min(\tfrac{1}{2}||w||^2 + \nu \sum_{i=1}^{n} \xi_i). \quad (3)$$

In this equation $\nu \in [0,1]$ is the regulator term or the Nu value (small $\nu$ allows margin constrains to be simply rejected whereas large $\nu$ makes the constrains hard to reject), for considering noise and overlapping of samples in dataset, $\xi$ parameter is used.
Nu-SVM is optimized using Bayesian Optimization which is a hyperparameter tuning approach which builds a probabilistic objective function using the Bayes Theorem and consequently selects a few candidate parameters to be evaluated with the main problem. Accuracy was chosen to maximize the classifier efficiency. Table summarizes the range of values that were inspected in this method. Optimal parameters are shown in Table 1 including: gamma=0.0098, number of iterations=163, and a Nu value of 0.4.

Table 1. Bayesian Optimization parameters

| Parameter | Value |
|---|---|
| Gamma | 0.0098 |
| Iterations | 176 |
| Nu | 0.4 |

2.7 Evaluation
In this study, 10-fold cross validation was used to evaluate the generalization ability of the model. In this technique, the dataset is divided into N approximately compliment subsets, N-1 folds regarded as training set and the remaining fold is test set. The process will repeat until every fold is once used as the test set. Finally, the average score of all N-folds presents the performance rate. One advantage of this technique is that the whole dataset is used for training and testing. Therefore, each sample appears in the training and testing sets and reduce the loss of information. Each performance component can be defined by equations below

$$Accuracy = TP + TN/TP + TN + FP + FN \quad (4)$$

$$Precision = TP/TP + FP \quad (5)$$

$$Recall = TP/TP + FN \quad (6)$$

$$F1(score) = 2 * (\frac{precision*recall}{precision+recall}) \quad (7)$$

2.8 Web based CAD implementation with Flask RESTful
In this work, a flask web service is built based on the trained model. The implementation is in a such way that a chest CT image is uploaded to the backend through an HTML form, making the image ready for preprocessing. Consequently, the proposed pipeline is evaluated on the image and the prediction is then shown to the user with the JSON format. To increase the prediction speed, models are preloaded before running the service. Figure 3 demonstrates this procedure in a web browser.

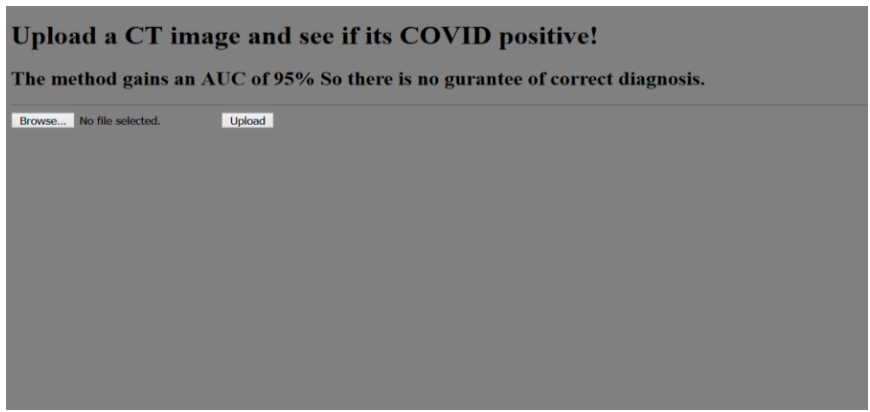

(a)

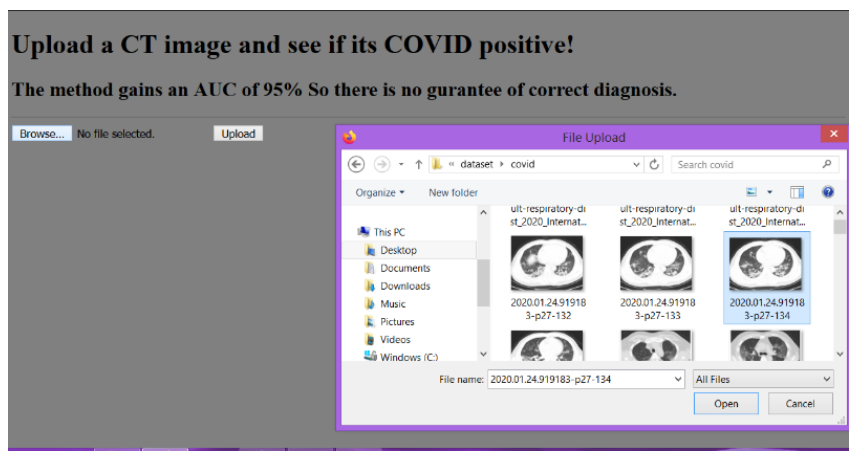

(b)

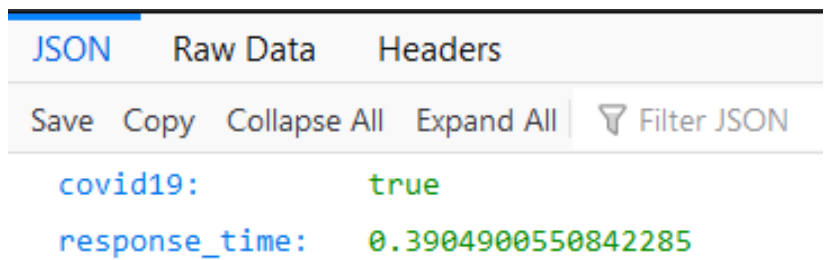

(c)

Figure 3. a) Representation of the proposed online detector web service b) adding a chest CT-image c) Results of the proposed online detector web service

## 3. Results

In this work, divergent Deep learning algorithms were inspected with the Nu-SVM. Figure 4. Represents the block diagram of entire process. All of the processing is done using python and a number of frameworks namely, Tensorflow, Keras, Scikit-learn. Common deep structures consisting of DenseNet121, ResNet50 V1&V2, InceptionV3 and MobileNet V1&V2 were used after resizing images phase to derive features from the images. The extracted maps were applied to SVM in order to classify samples in two categories.

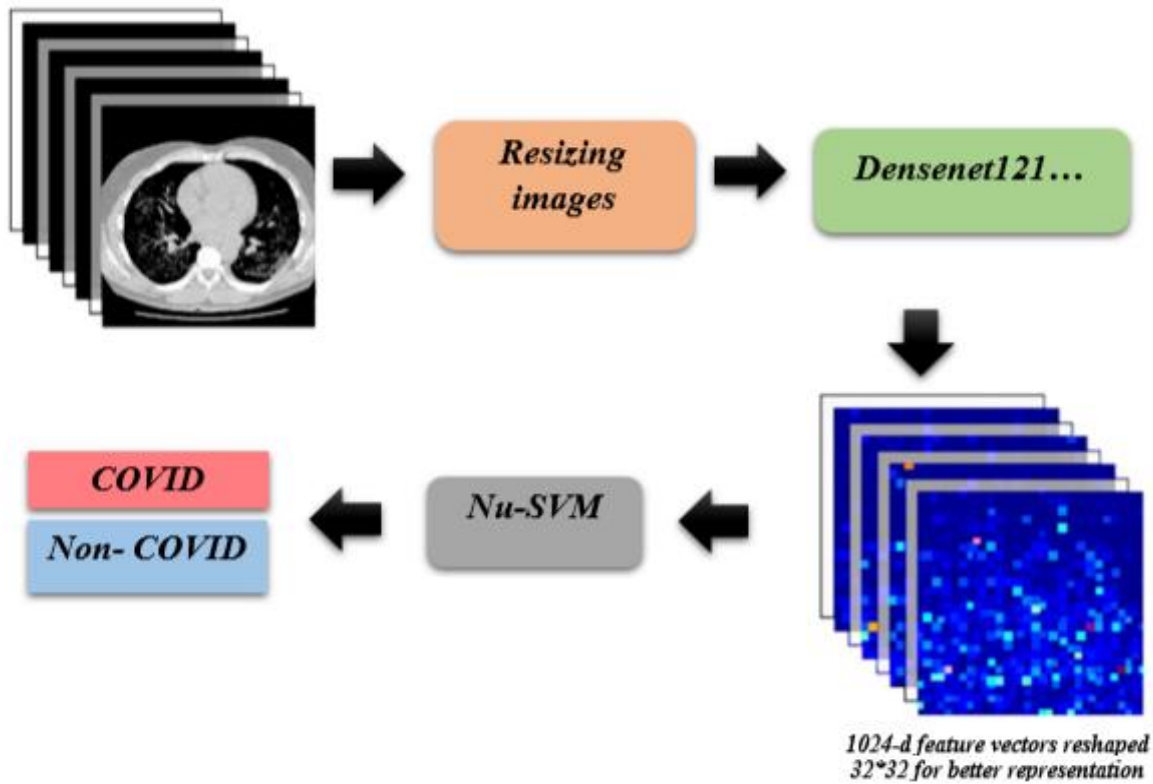

Figure 4. Demonstrations of automatic COVID-19 detection

Initially feature maps were extracted from the model, resulting in a 1024-dimensional feature map which was is reshaped for a better representation and a sample is shown in (Figure 5).

The performances achieved for 10-fold cross validation includes of accuracy, recall, precision, f1-score and AUC for each deep learning methods are demonstrated in Table 2.

Figure 6 displays the fold variations of each performance metric over the 10-folds for the DenseNet121 model. As it can be seen, 9.2% of the COVID-19 misclassified as healthy controls and 10.24% of the normal participants incorrectly categorized as respiratory patients.

Table 2. Performance values achieved by testing the model

|  | Accuracy | Recall | Precision | F1-Score | AUC | Output size |
|---|---|---|---|---|---|---|
| DenseNet121 | **90.61(±5.4)** | **90.80(±5.3)** | **89.76(±3.7)** | **90.13(±5.4)** | **95.05(±2.9)** | 1024 |
| InceptionV3 | 81.63(±3.7) | 80.08(±5.3) | 80.07(±6.3) | 80.49(±3.4) | 87.78(±3.1) | 2048 |
| ResNet50V2 | 84.03(±4.6) | 82.23(±7.5) | 83.83(±5.7) | 82.76(±4.9) | 84.49(±3.7) | 2048 |
| ResNet50V1 | 73.72(±3.9) | 73.94(±6.5) | 71.85(±3.6) | 72.46(±3.6) | 79.18(±5.3) | 2048 |
| MobileNetV1 | 89.14 (±2.6) | 88.53(±4.2) | 88.64(±5.1) | 88.43(±2.7) | 94.15(±1.9) | 1024 |
| MobileNetV2 | 85.52(±3.7) | 87.66(±4.4) | 82.84(±5.5) | 85.03(±3.6) | 92.78(±2.7) | 1024 |

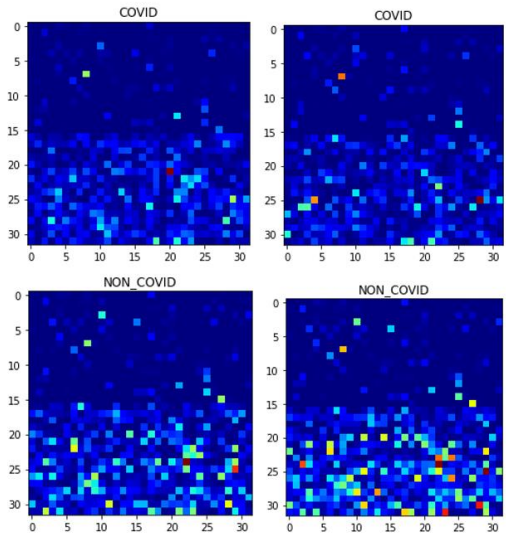

Figure 5. Illustrations of feature maps extracted by DenseNet121

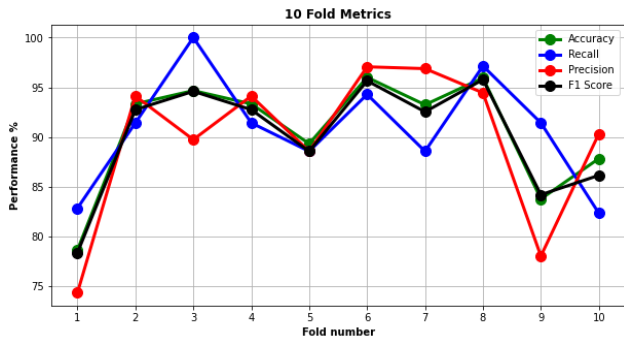

Figure 6. Performance measures variations over 10-fold for the DenseNet-SVM model

Another model efficiency measure is the Receiver operating characteristics (ROC) curve which is constructed by plotting the true positive rates at the different threshold levels versus the false positive rates. Figure 7 shows the ROC plot for the proposed method. As it can be seen, the DenseNet121 model is consistent over all folds, resulting in an Area Under Curve of 95%.

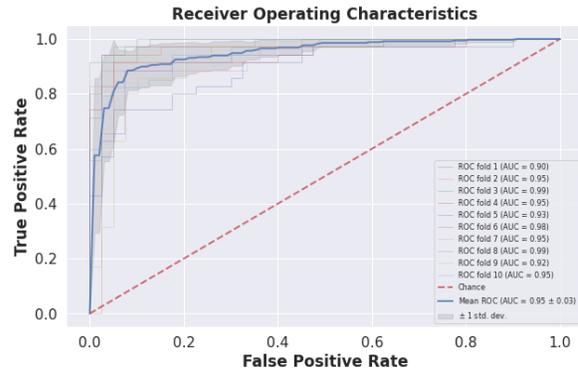

Figure 7. ROC Curve of proposed DenseNet-SVM system

MobileNetV1&V2 in comparison with other models yield considerable performance rates with 89.14%, 85.52% accuracy respectively. The 94.15% and 92.78% AUC of these methods also demonstrates high diagnostic ability of binary classifiers. (Table2)

4. Discussion

Deep learning techniques have made considerable advances in the field of automated diagnostics systems and prevailed on the limitations of conventional machine learning algorithms. Fang et al [15] compared the CT-scan and RT-PCR for COVID-19 detection. Their results demonstrate that sensitivity for CT-scan based detection with 98% was considerably higher than the RT-PCR technique with 71% which is a motivation to enhance CT based techniques. In our method a combined DenseNet-Nu-SVM system was proposed for processing CT-scan images. With database consisting of (349 COVID and 397 Non-COVID) 746 patients, which is more dependable compared to most of the studies which had less than a hundred COVID patients [2-5]. DenseNet121 was used to reduce parameter space by extracting local features and Nu-SVM was used for classification According to Table 2, the implemented system yielded 90.61 percentage accuracy and 90.80 percentage sensitivity which is superior compared to the performance rates of other popular architectures. 95.05% AUC was also achieved which demonstrates the hybrid system monitoring ability. In addition, the suggested diagnostic approach was developed as a web-service program which simply gives the results of COVID diagnosis to the user after uploading a chest CT-scan image. The advantage of the proposed methodology is a reasonable accuracy as well as online usability (figure 8). A disadvantage would be the small dataset size in which, since deep learning models training process requires a considerable number of samples and small datasets lead to overfitting, combining non biased DNN derived features with Nu-SVM could overcome this issue and perform well with classifying previously unseen data.

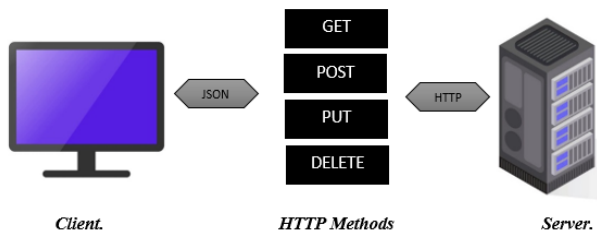

Figure 8. Diagram of the proposed web service

## 5. Conclusion

In this research a pipeline was introduced to detect COVID-19 easily and accurately. Different pretrained models were compare and the mot significant approach was deployed as a RESTful flask web service with the aim of online COVID-19 detection. Metrics included an accuracy of 90.61%, 90.80% sensitivity and 95.05 AUC. The experimental results demonstrate that the proposed diagnostic approach has the ability of monitoring COVID-19. Moreover, it may be possible to build more reliable models with this approach using larger databases and use them anywhere in the world.

References
[1] https://www.who.int/emergencies/diseases/novel-coronavirus-2019/situation-reports
[2] Shuai Wang1, Bo Kang2,3, Jinlu Ma4, Xianjun Zeng5, Mingming Xiao1, Jia Guo3, Mengjiao Cai4, Jingyi Yang4, Yaodong Li6, Xiangfei Meng2, Bo Xu1, deep learning algorithm using CT images to screen for Corona Virus Disease (COVID-19).
[3] COVID-CT-Dataset: A CT Scan Dataset about COVID-19 Jinyu Zhao, UC San Diego, Yichen Zhang, UC San Diego, Xuehai He, UC San Diego, Pengtao Xii, UC San Diego, Petuum Inc.
[4] Linda Wang and Alexander Wong, COVID-Net: A Tailored Deep Convolutional Neural Network Design for Detection of COVID-19 Cases from Chest Radiography Images
[5] Xiaowei Xu1, MD; Xiangao Jiang2, MD, Chunlian Ma3, MD; Peng Du4; Xukun Li4; Shuangzhi Lv5, MD; Liang Yu1, MD; Yanfei Chen1, MD; Junwei Su1, MD; Guanjing Lang1, MD; Yongtao Li1, MD; Hong Zhao1, MD; Kaijin Xu1, PhD MD; Lingxiang Ruan5, MD; Wei Wu1, PhD MD, Deep Learning System to Screen Coronavirus Disease 2019 Pneumonia
[6] COVID-CT-Dataset: A CT Scan Dataset about COVID-19 Jinyu Zhao, UC San Diego, Yichen Zhang, UC San Diego, Xuehai He, UC San Diego, Pengtao Xii, UC San Diego, Petuum Inc. https://github.com/UCSD-AI4H/COVID-CT.

[7] Mengye Ren∗, Renjie Liao∗, Raquel Urtasun, Fabian H. Sinz, Richard S. Zemel NORMALIZING THE NORMALIZERS: COMPARING AND EXTENDING NETWORK NORMALIZATION SCHEMES

[8] G. Huang, Z. Liu, L. Van Der Maaten and K. Q. Weinberger, "Densely Connected Convolutional Networks," 2017 IEEE Conference on Computer Vision and Pattern Recognition (CVPR), Honolulu, HI, 2017, pp.2261-2269.doi: 10.1109/CVPR.2017.243

[9] K. He, X. Zhang, S. Ren and J. Sun, "Deep Residual Learning for Image Recognition," 2016 IEEE Conference on Computer Vision and Pattern Recognition (CVPR), Las Vegas,NV, 2016,pp.770-778. doi: 10.1109/CVPR.2016.90

[10] Gustav Larsson, Michael Maire, Gregory Shakhnarovich, FRACTALNET:ULTRA-DEEP NEURAL NETWORKS WITHOUT RESIDUALS

[11] Christian Szegedy, Vincent Vanhoucke, Sergey Ioffe, Jonathon Shlens ,Rethinking the Inception Architecture for Computer Vision

[12] Christian Szegedy, Pierre Sermanet, Wei Liu, Yangqing Jiao, Dumitru Erhan, Scott Reed, Dragomir Angelou, Andrew Rabinovich,Vincent Vanhoucke. Going deeper with convolutions

[13] MobileNets: Efficient Convolutional Neural Networks for Mobile Vision Applications Andrew G. Howard Menglong Zhu Bo Chen Dmitry Kalenichenko Weijun Wang Tobias Weyand Marco Andreetto Hartwig Adam

[14] CORINNA CORTES, VLADIMIR VAPNIK, Support-Vector Networks

[15] Yicheng Fang, MD1 Huangqi Zhang, MD1 Jicheng Xie, MD1 Minjie Lin, MD1 Lingjun Ying, MD2 Peipei Pang, MD3 Wenbin Ji, MD1, Sensitivity of Chest CT for COVID-19: Comparison to RT-PCR